\documentclass{jpsj2}
\title{A Symmetry Property of Momentum Distribution Functions in
the Nonequilibrium Steady State of Lattice Thermal Conduction}

\author{
Akira \textsc{Ueda}$^{1}$
\thanks{E-mail: ueda@ms.osakafu-u.ac.jp}
and Shinji \textsc{Takesue}$^{2}$
\thanks{E-mail: takesue@phys.h.kyoto-u.ac.jp}
}

\inst{
$^{1}$Department of Mathematical Sciences, Osaka Prefecture University, Sakai,
699-8531, Japan \\
$^{2}$Department of Physics, Kyoto University, Sakyo-ku, Kyoto, 
606-8501, Japan \\
}

\abst{
We study a symmetry property of momentum distribution functions in the
steady state of heat conduction.  When the equation of motion is symmetric
under change of signs for all dynamical variables, the distribution function 
is also symmetric.  This symmetry can be broken by introduction of an
asymmetric term in the interaction potential or the on-site potential, or
employing the thermal walls as heat reservoirs.
We numerically find differences of behavior of the models with and without 
the on-site potential.
}

\kword{Lattice dynamical systems, heat conduction, nonequilibrium steady
state, Maxwellian distribution, symmetry}

\begin{document}
\maketitle
\section{Introduction} 
Recently, considerable progress has been made in statistical mechanics of 
nonequilibrium steady states beyond the linear-response regime.  
First, new formulas such as the fluctuation theorem\cite{Evans} and the 
Jarzynski equality\cite{Jarzynski} have been found.  These formulas can be used
as a basis for arguments on transport phenomena in small systems, such as 
energy transfer in motor proteins or heat and electric transport in nano-scale
devices.  Second, Sekimoto proposed energetics on Langevin systems, 
namely systems
in the scale where fluctuations become important\cite{Sekimoto}.
It was further developed and some interesting results were obtained\cite{Sasa}
Lastly, some exactly solvable models were found\cite{ASEP,Sasamoto}, 
which give us important insights to the problem.

Lattice thermal conduction is a typical example of the nonequilibrium steady
states.  It may be more fundamental than stochastic systems 
because it is described by a Hamiltonian.  
Study of thermal conductivity has a long history, but many important results 
were obtained only recently.  It was established already in 1960s that 
integrable systems do not have temperature gradient 
and heat flux is proportional to not temperature gradient but temperature 
difference between the ends.  
For example, heat conduction in the harmonic chain was solved in 
1958\cite{Rieder}. 
However, it was found in 1997 that the thermal conductivity in the
Fermi-Pasta-Ulam (FPU) chain diverges like 
$\kappa\sim N^{\alpha}$ ($\alpha\simeq 0.37$) as the system size $N$ is
increased\cite{Lepri1}.  
Since then, a lot of researches were carried out and it was clarified 
that the divergence is caused by a long-time tail in current
autocorrelation function, which is originated from momentum
conservation\cite{Hatano,Lepri2,Narayan}. 
On the contrary, 
if the system includes a potential force from substrate the thermal
conductivity is convergent.  The $\phi^4$ system\cite{Aoki}, 
the ding-a-ling model\cite{Mimnagh} and
the ding-dong model\cite{Sano} belong to the class of systems.  
However, we must mention that exceptions exist for this rule and 
that proposed theories have some discrepancy
concerning the value of exponent $\alpha$\cite{Lepri3,Narayan}.  
Thus, we are still far from complete understanding.

As we see in the above, 
the study of lattice thermal conduction has been restricted to the Fourier
heat law and existence or nonexistence of thermal conductivity. 
In fact, 
there are few studies on distribution functions that describe the 
nonequilibrium steady state of heat conduction.  
To construct nonequilibrium statistical mechanics,
however, properties of the distribution function should be studied.  
Thus, in this paper we investigate characteristics of the nonequilibrium 
steady states.

Analogously with the Gibbs ensemble in equilibrium statistical mechanics, 
there
should be a distribution function 
on the phase space that describes a nonequilibrium steady state.  
However, the phase space is high-dimensional and numerically 
intractable.  So, we focus on momentum distribution of a single particle.  
In equilibrium, it is a Maxwellian distribution at some temperature.  
We study how it deviates from the Maxwellian distribution in nonequlibrium 
steady states.

In particular, we focus on a symmetry property.  In equilibrium systems, the
momentum distribution does not depend on the potentials.  On the other hand,
symmetry of the potentials can affect the momentum distribution in
nonequilibrium states.  We numerically investigate what happens if the
symmetry is broken by introducing a small asymmetric term into the on-site
potential or the interaction potential or employing the thermal wall as the
heat reservoir.  As a result, differences are found in the behavior of the
models with and without on-site potentials.  This may be relevant with the
reported behavior of heat conduction and might be useful for full
understanding of the problem.  

In fact, Aoki and Kusnezov\cite{AK2004} discussed similar deviation from the 
Maxwellian distribution and derived a scaling form for the fourth-order
cumulants.  However, they were limited to symmetric systems and symmetric
deviations.  In this paer, we extend to asymmetric models and asymmetric
deviations.

In Section 2, we describe the models and heat reservoirs employed in our
simulations.  In Section 3, we demonstrate a relation between the symmetry
in the equation of motion and that in the momentum distribution functions.
In Section 4, we show our numerical results when a weak asymmetry is
introduced to a symmetric model.  Section 5 is devoted to summary and
discussion.

\section{Models and heat reservoirs}
We consider one-dimensional systems composed of $N$ particles of unit mass 
with a Hamiltonian of the form
\begin{equation}
 H(\{q_n\},\{p_n\})=
\sum_{n=1}^{N}\left[\frac{p_n^2}{2}+U(q_n)\right]
+\sum_{n=0}^{N}V(q_{n+1}-q_{n}),
\label{models}
\end{equation} 
where $q_n$ and $p_n$ are the displacement and the momentum of 
particle $n\in\{1,2,\dots,N\}$, 
$U(q)$ is an on-site potential, and $V(q)$ is an
interaction potential between nearest-neighbor particles.  
We impose the fixed boundary condition, which is represented
by setting $q_{0}=q_{N+1}=0$ in Eq.\/ (\ref{models}).
Various models are generated by varying $U$ and $V$.  Typical examples
are (a) the Fermi-Pasta-Ulam (FPU) model ($U(x)=0$, and
$V(x)=\frac{x^2}{2}+\frac{x^4}{4}$), (b) the $\phi^4$ model
($U(x)=\frac{x^2}{2}+\frac{x^4}{4}$, and $V(x)=\frac{x^2}{2}$), 
and (c) the Toda model ($U(x)=0$ and $V(x)=\exp(x)-x$).

The particles at the ends of the chain are in contact with two heat 
reservoirs: one at temperature $T_L$ on particle 1 and the other at 
temperature $T_R$ on particle $N$.  If $T_L=T_R$, the system goes to
equilibrium, and if $T_L\ne T_R$, heat conduction occurs.
Thus, for $2\le n\le N-1$, the equations of motions
are given as
\begin{equation}
\dot{q}_n=p_n,\quad \dot{p}_n=-U'(q_n)+V'(q_{n+1}-q_{n})-V'(q_{n}-q_{n-1}),
\label{eqofmotion}
\end{equation}
while those for particles $1$ and $N$ are modified from the Hamiltonian form.
In this paper, we consider the following three kinds of reservoirs to study
what differences are generated by the types of heat reservoirs.
The first is the Langevin reservoir that is given by adding a dissipation term
and fluctuating force to the equation of the motion as
\begin{equation}
 \dot{p}_1=-U'(q_1)+V'(-q_1)-V'(q_1-q_2)-\gamma p_1+\xi_L(t)
\label{Langevin1}
\end{equation}
and
\begin{equation}
 \dot{p}_N=-U'(q_N)+V'(q_{N-1}-q_N)-V'(q_{N})-\gamma p_N+\xi_R(t).
\label{LangevinN}
\end{equation}
In the above equations, $\xi_L(t)$ and $\xi_R(t)$ are Gaussian white noise 
with correlation
\[
 \langle \xi_\alpha(t)\xi_{\beta}(t')\rangle =2\gamma k_BT_\alpha\delta_{\alpha\beta}\delta(t-t')
\]
where $\alpha, \beta=L$ or $R$.  The second is the Nose-Hoover
reservoir\cite{Nose}, 
which is a kind of deterministic reservoir and widely used in
the literature.    
In this reservoir, the equations of motion for particles $1$ and $N$ are 
modified as  
\begin{equation}
\dot{p}_1=-U'(q_1)+V'(q_2-q_1)-V'(q_1)-\xi_Lp_1,\quad
\dot{\xi_L}=\frac{1}{\Theta}\biggl(\frac{p_1^2}{T_L}-1\biggr),
\label{NH1}
\end{equation}
and 
\begin{equation}
\dot{p}_N=-U'(q_N)+V'(-q_N)-V'(q_N-q_{N-1})-\xi_Rp_N,\quad
\dot{\xi_R}
=\frac{1}{\Theta}\biggl(\frac{p_N^2}{T_R}-1\biggr),
\label{NHN}
\end{equation}
and time evolution of $\xi_L$ and $\xi_R$ are given by
\begin{equation}
 \dot{\xi_L}=\frac{1}{\Theta}\biggl(\frac{p_1^2}{T_L}-1\biggr)
\label{NH1xi}
\end{equation}
and
\begin{equation}
 \dot{\xi_R}=\frac{1}{\Theta}\biggl(\frac{p_N^2}{T_R}-1\biggr),
\label{NHNxi}
\end{equation}
where, $\Theta$ is a constant which means response time of the thermostat.
The last is the thermal wall\cite{thermalwall}.  When a particle hit the
thermal wall, it is reflected with a momentum chosen according to
the probability distribution
\begin{equation}
 f(p)=\frac{|p|}{T}\exp\left(-\frac{p^2}{2T}\right)
\end{equation}
where $T$ denotes the temperature of the wall.  For our systems, we place a
thermal wall at temperature $T_L$ on the left hand of particle $1$ and 
the other wall at temperature $T_R$ on the right hand of particle $N$.

The three models mentioned in Section 1 are representatives of the types of
heat conduction behavior.
Namely, the FPU model represents the models where the thermal
conductivity diverges in the thermodynamic limit due to a long-time tail
in the autocorrelation function of heat flux.  The $\phi^4$ model
represents the class of models where the thermal conductivity converges
as the system size is increased.  The Toda model represents the
integrable models where temperature gradients are not formed and heat
flux is proportional to temperature differences between the reservoirs.

\section{Deviation from the Maxwellian distribution}
We carriy out numerical simulations on the three models with Langevin heat
reservoirs at $T_L=2.0$ and $T_R=1.0$ and calculate the single-particle 
momentum distribution functions $P_n(p)$ ($n=1,\dots,N$) as follows.  
The deterministic part of the equation of motion is solved with a mixed use 
of the sixth-order symplectic method and the fourth-order Runge-Kutta method 
with a single time step $\Delta t=0.01$.
After computing the deterministic part, 
we add a Gaussian random noise with 
variance $2\gamma T_L\Delta t$ to the momentum of
particle $1$, and one with variance $2\gamma T_R\Delta t$ to the momentum of
particle $N$. 
To compute the distribution functions, simulation was done during 
$10^{10}$-steps after the system reached a steady state.
Figure 1 shows the difference between the computed momentum distribution
function and the Maxwellian distribution function 
\begin{equation}
 P_{\mathrm{M}}(p;T_n)=\frac{1}{\sqrt{2\pi T_n}}
\exp\left(-\frac{p^2}{2T_n}\right),
\end{equation}
where temperature $T_n$ is determined via average of kinetic energy, namely
$T_n=\langle p_n^2\rangle$.

We find that the deviation is symmetric in the FPU model and the $\phi^4$ 
model, while it is asymmetric in the Toda chain.  This is
because the equations of motion are symmetric under 
the transformation of changing signs of the variables for the former two
models, but not for the Toda chain.  
In general, the equation of motion (\ref{eqofmotion}) is invariant under the
transformation of changing the signs of the variables
$(q_1,p_1;\dots;q_N,p_N)\rightarrow(-q_1,-p_1;\dots;-q_N,-p_N)$
if both the on-site potential $U(q)$ and the interaction 
potential $V(q)$ are even functions, i.e., $U(-q)=U(q)$ and $V(-q)=V(q)$.
We here note that there is no physical reason that the interaction potential 
must be even.  Denoting lattice constant by $c$,
equality $V(-q)=V(q)$ means that 
two particles with distances $c+q$ and those with distance $c-q$ have 
the same interaction energy, which is not expected in general.

Moreover, the Langevin equations for particles 1 and $N$, (\ref{Langevin1})
and (\ref{LangevinN}), are also invariant under the transformation if the
signs of the noise terms $\xi_L(t)$ and $\xi_R(t)$ are also changed.  The
last operation does not change the statistics of the noise terms.
Similarly, Nose-Hoover reservoir is also symmetric under operation  
$(q_1,p_1;\dots;q_N,p_N;\xi_L,\xi_R)\rightarrow(-q_1,-p_1;\dots;-q_N,-p_N;\xi_L,\xi_R)$.
On the other hand, since the thermal wall introduces an asymmetric
potential, it breaks the symmetry.

\begin{figure}[tb]
\begin{center}
\begin{tabular}{cc}
\includegraphics[width=7cm]{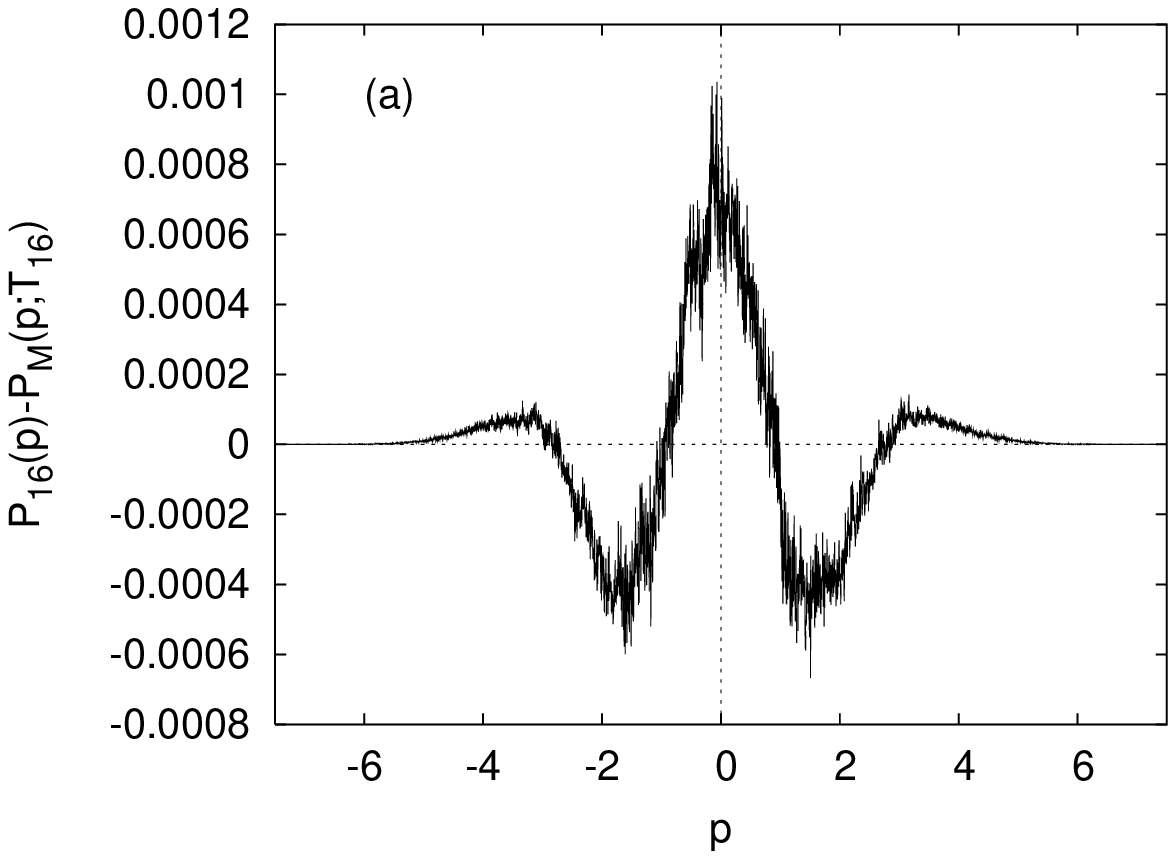}&
\includegraphics[width=7cm]{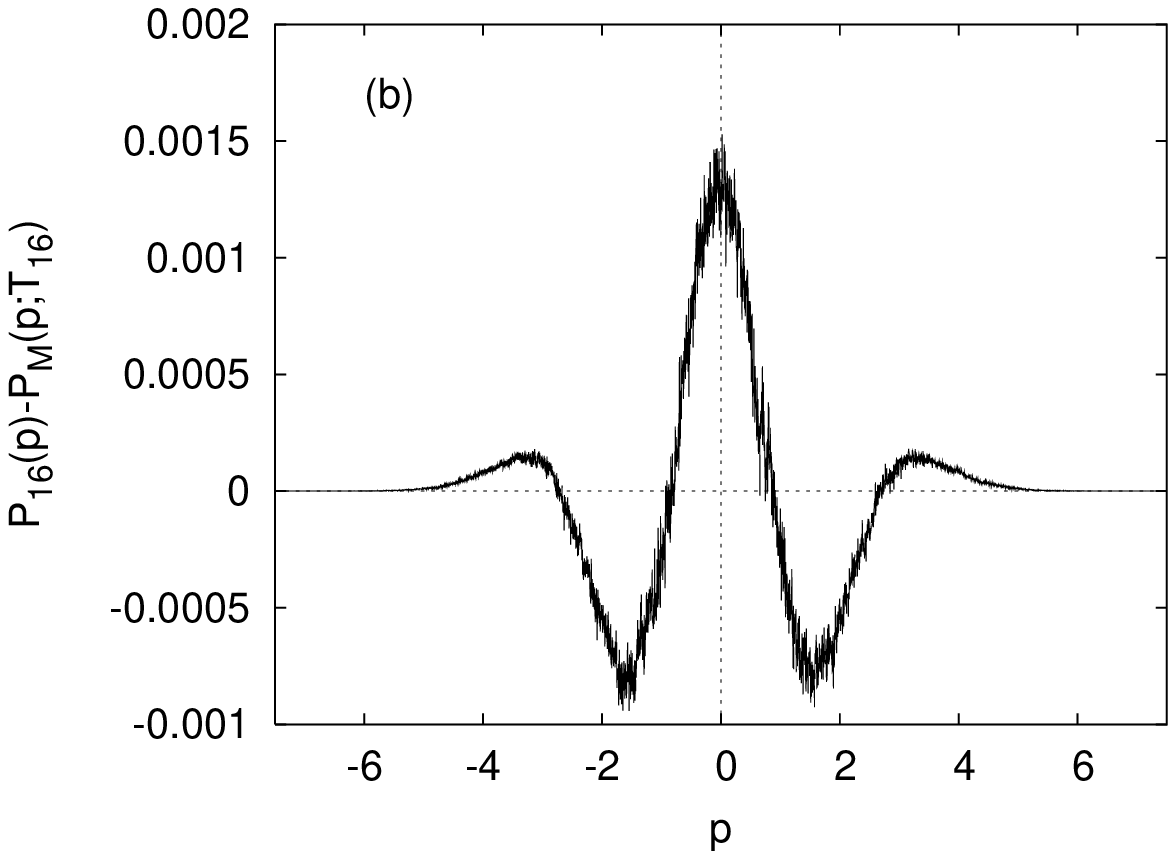}\\
\includegraphics[width=7cm]{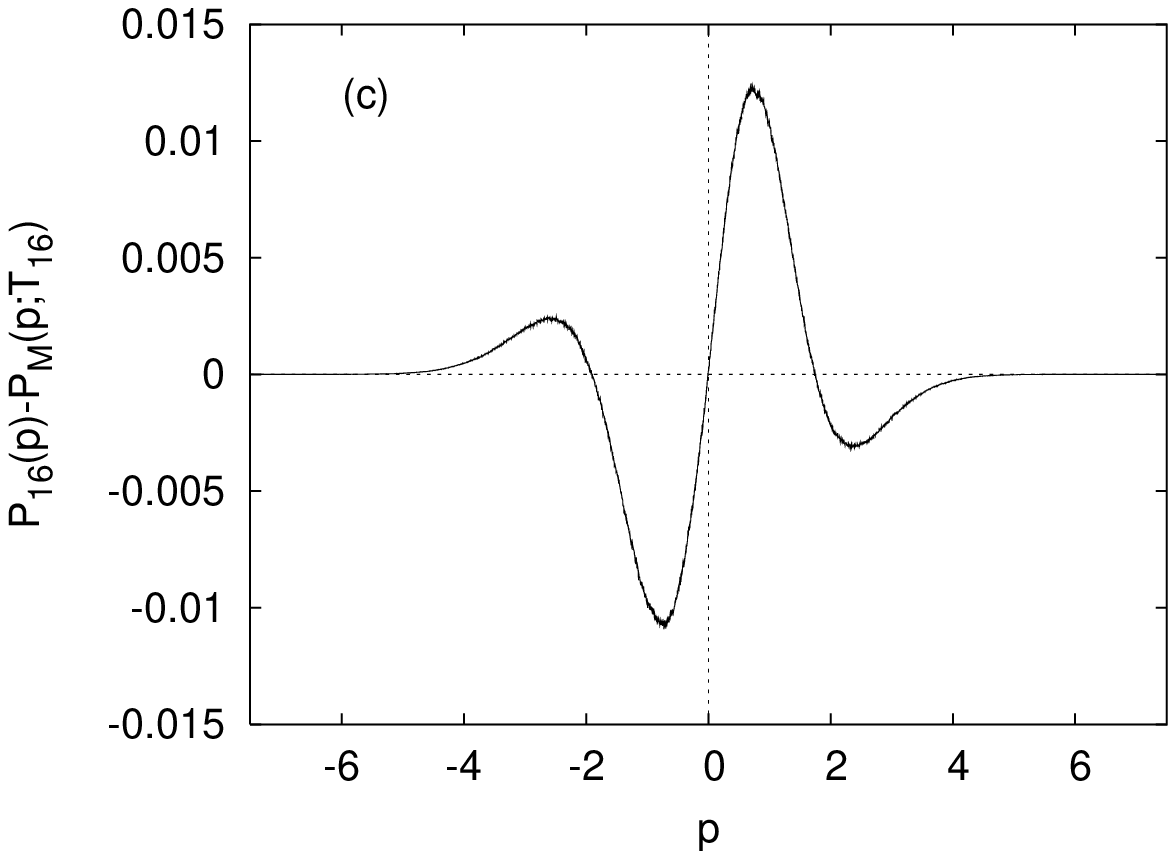} & 
\end{tabular}
\end{center}
\caption{Deviation from the Maxwellian distribution for (a) the FPU model,
 (b) the $\phi^4$ model, and (c) the Toda model.  Here the system size is $N=32$ and
the temperature values of the heat reservoirs are given as $T_L=2.0$ and $T_R=1.0$. $10^{10}$ 
iterations are carried out for (a), (b) and (c).}
\label{f1}
\end{figure}

The symmetric deviation seen in the FPU model and the $\phi^4$ model is in
contrast with the antisymmetric deviation seen in particle systems described 
by the Boltzmann equation\cite{Kim}.   
In the latter case, the expectation value of heat current is represented
as $\frac{1}{2}\int\boldsymbol{p}p^2f(\boldsymbol{p})d\boldsymbol{p}$ with use
of the momentum distribution function $f(\boldsymbol{p})$.  
Clearly, it vanishes if $f(\boldsymbol{p})$ is symmetric.
Thus, antisymmetric deviation must exist.
On the other hand, in our lattice models the expectation value of 
heat current is represented as 
\[
 -\int\left(p_n+p_{n+1}\right)V'(q_{n+1}-q_{n})P_{n,n+1}(q_{n},p_{n},q_{n+1},p_{n+1})
 dq_ndp_ndq_{n+1}dp_{n+1},
\]
where $P_{n,n+1}(q_{n},p_{n},q_{n+1},p_{n+1})$ denotes a two-body
distribution function in the steady state.  Because $V'(q)$ is antisymmetric
when $V(q)$ is symmetric, the integral does not vanish if 
$P_{n,n+1}(q_{n},p_{n},q_{n+1},p_{n+1})$ is symmetric.
This also indicates the existence of correlation.
Namely, if $P_{n,n+1}(q_{n},p_{n},q_{n+1},p_{n+1})$ is decomposed into a 
product form
$P_{n,n+1}(q_{n},p_{n},q_{n+1},p_{n+1})
=P_{n,n+1}^{(q)}(q_{n},q_{n+1})P_{n,n+1}^{(p)}(p_{n},p_{n+1})$
as in the equilibrium case and $P_{n,n+1}^{(q)}$ and/or $P_{n,n+1}^{(p)}$ is
symmetric, the heat current vanishes. 
Thus, to produce nonzero heat current some correlation must exist between
momentum and position.  The symmetric deviation in the momentum distribution
function is a reflection of such correlation.

\section{Properties of asymmetric deviation}

In this section, we examine how the momentum distribution functions are
affected if the symmetry is broken by small modifications to a symmetric
system.  As we noted in the Introduction, there are three kinds of such
modifications, which we describe in the following.
 
\begin{figure}[htbp]
\begin{center}
\includegraphics[width=7cm]{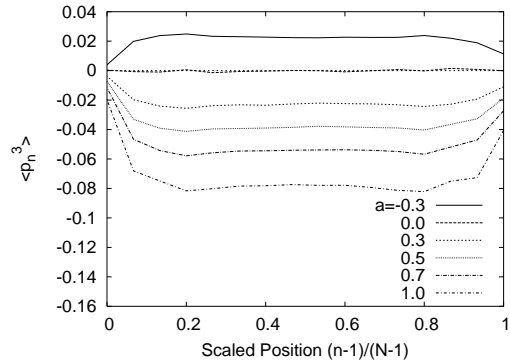}
\end{center}
\caption{\small The third-order moments $\langle p_n^3\rangle$ in the
   system with asymmetric interaction potential (\ref{asym_pot}) and 
on-site potential $U(q)=0$. 
The system size is $N=16$, parameter values are $a=-0.3$, $0$, $0.3$, $0.5$, 
$0.7$, $1.0$. Langevin reservoirs at $T_L=2.0$ and $T_R=1.0$ are used.}
\label{f2}
\end{figure}

\begin{figure}[htbp]
\begin{center}
\includegraphics[width=7cm]{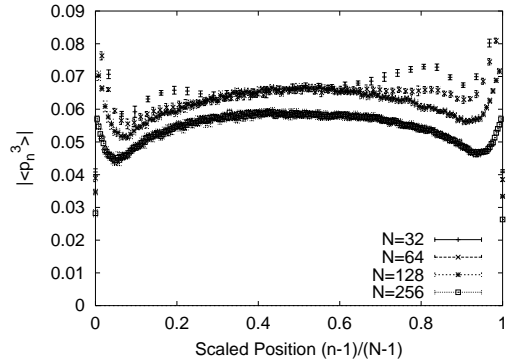}
\end{center}
\caption{\small The third-order moments $|\langle p_n^3\rangle |$ in various
 sizes of the system having the asymmetric potential (\ref{asym_pot}) with
$a=1$.  The reservoir temperatures are given as $T_L=2.0$ and $T_R=1.0$.
}
    \label{fpu_size}
\end{figure}

\begin{figure}[htbp]
\begin{center}
 \includegraphics[width=7cm]{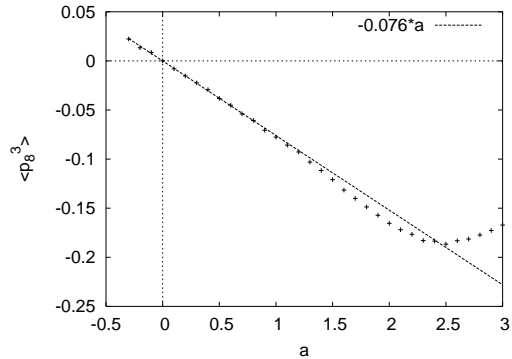}
\end{center}
\caption{\small
Variation of  $\langle p_8^3\rangle$ as the parameter $a$ is
 varied.  The system is the same as in Fig.\/ \ref{f2}}
\label{f2x} 
\end{figure}

The first is the case of modifying the interaction potential. 
Let us consider a system with the following interaction potential,
\begin{equation}
V(q)=\frac{1}{2}q^2+\frac{a}{3}q^3+\frac{1}{4}q^4,
\label{asym_pot}
\end{equation}
with a constant $a$ representing the magnitude of asymmetry. 
The on-site potential $U(q)$ is set to be zero.  
We carry out numerical simulations of the system of size $N=16$ and 
calculate the third-order moments 
\begin{equation}
\langle
 p_n^3\rangle=\int^{+\infty}_{-\infty}p^3P_n(p)dp\hspace{10pt}(n=1,...,N).
\end{equation}
The results are shown in Fig.\/ \ref{f2}, which
indicates that asymmetric deviations from the Maxwellian distribution 
emerge in the whole system.  Namely, the moments are 
in the same order of magnitude except a few particles near the ends. 
This behavior is not changed if the system size is varied.  
See Fig.\/ \ref{fpu_size}.
Although there appear relatively large variations near both the ends,  
changes are smooth in the middle of the system.
Figure \ref{f2x} shows that the third-order moment changes in proportion to
$a$ when $a$ is small.  But the tendency changes around $a\simeq 2$,
where the moment have an extremum. This is because the stability of $q=0$ is lost at $a=2$ and two stable points with $q\ne 0$ appear beyond $a=2$.  Then, $q$ loses the meaning of displacement from the stable point in the latter case. 

We check this behavior with some conditions modified. 
One is the introduction of on-site potential.  When we use the nonlinear
on-site potential 
\[
 U(q)=\frac{1}{2}q^2+\frac{1}{4}q^4,
\]
the third-order moments become almost constant except for the end
particles and this behavior does not change with the system size.
The next is the use of Nose-Hoover reservoirs instead of Langevin reservoirs.
In this case, the magnitude of the third-order moments is larger than the
Langevin case and their spatial variation is rather smooth.  No noticeable
peaks are formed if the system size is changed.  The next is the variation of
the boundary temperatures.  When the temperatures of the reservoirs are given
as $T_L=60$ and $T_R=30$, 
the overall behavior is the same as the low temperature case, only
magnifying the absolute values of the moments.
The last is insert of the fifth-order term instead of the third-order term
into the interaction potential.  The use of the fifth-order term leads to
little difference from the case of third-order term. 
In all cases, the overall appearance of the third-order moments are observed.

\begin{figure}[htbp]
\begin{center}
\begin{tabular}{cc}
\includegraphics[width=7cm]{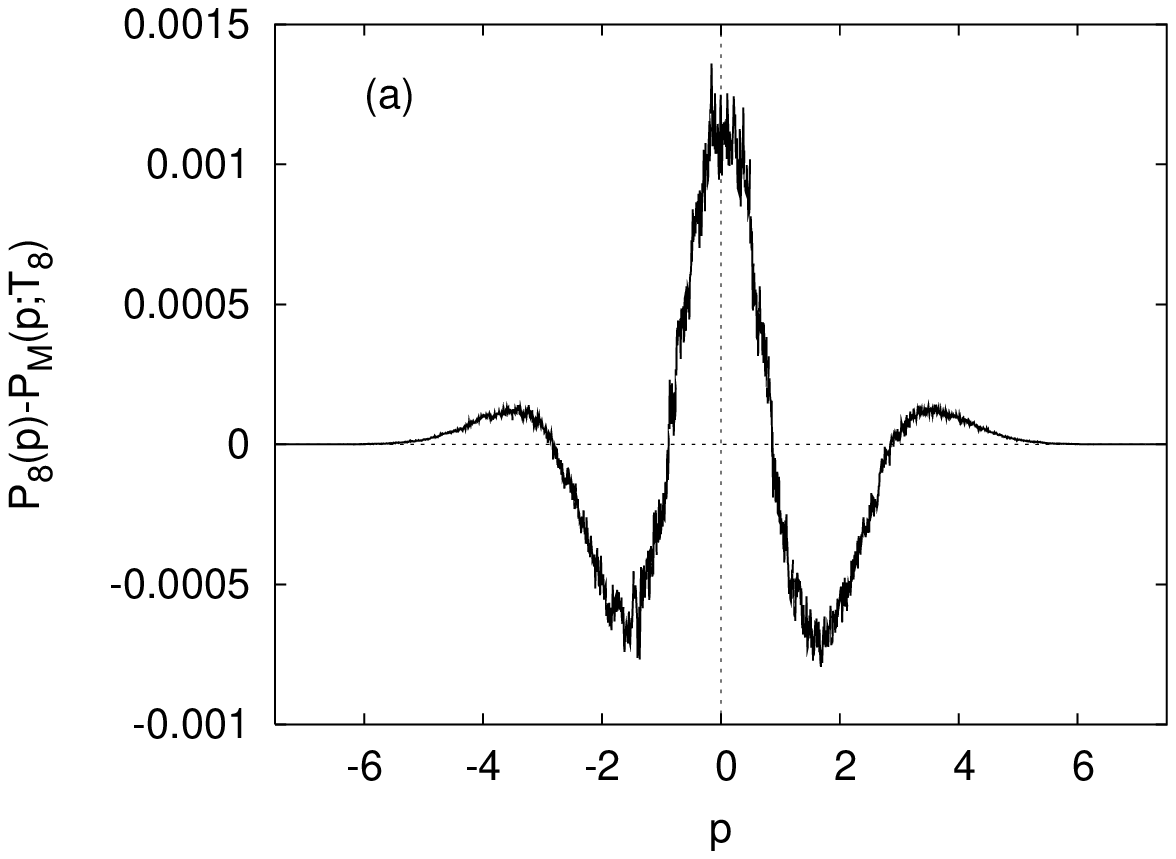}&
\includegraphics[width=7cm]{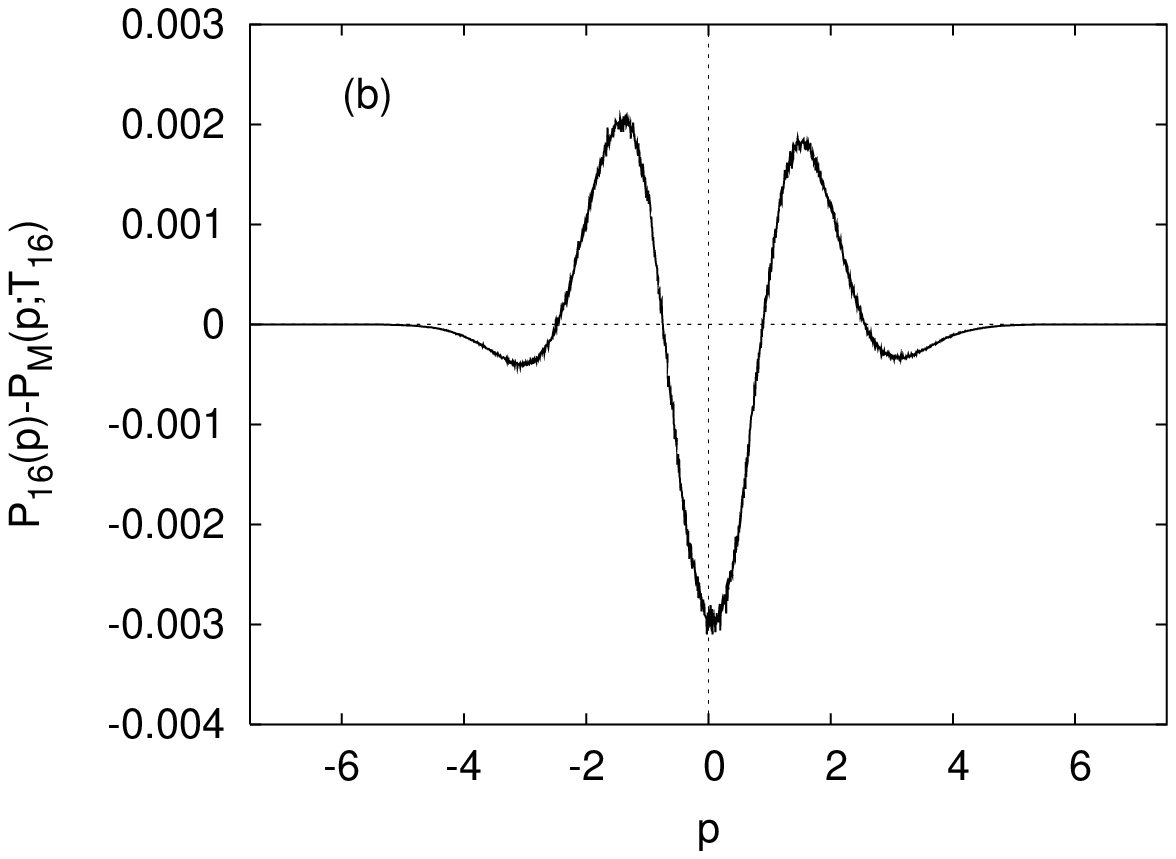}
\end{tabular}
\end{center}
\caption{\small Deviations from the Maxwellian distribution for (a) particle
$8$ and (b) particle $16$ in the model with the asymmetric on-site 
potential $U(q)=\frac{1}{2}q^2+\frac{1}{3}q^3+\frac{1}{4}q^4$ and the
harmonic interaction potential $V(q)=\frac{1}{2}q^2$.  The system size is
 $N=16$ and Langevin reservoirs at $T_L=2.0$ and $T_R=1.0$ are used. $10^{10}$ 
iterations are carried out for each of them.}
    \label{f3}
\end{figure}

\begin{figure}[htbp]
\begin{center}
\includegraphics[width=7cm]{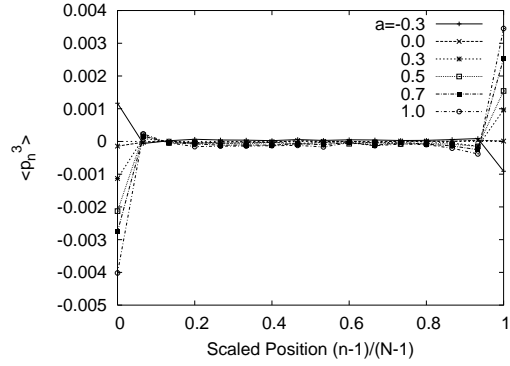}
\end{center}
\caption{\small The third-order moments $\langle p_n^3\rangle$  
in the system with the asymmetric on-site potential 
$U(q)=\frac{q^2}{2}+a\frac{q^3}{3}+\frac{q^4}{4}$ and the harmonic 
interaction potential $V(q)=\frac{q^2}{2}$.  The system size is $N=16$, 
Langevin reservoirs at $T_L=2.0$ and $T_R=1.0$ are used, and 
parameter values are $a=-0.3$, $0$, $0.3$, $0.5$, $0.7$, and $1.0$.}
    \label{f4}
\end{figure}

\begin{figure}[htbp]
 \begin{center}
  \includegraphics[width=7cm]{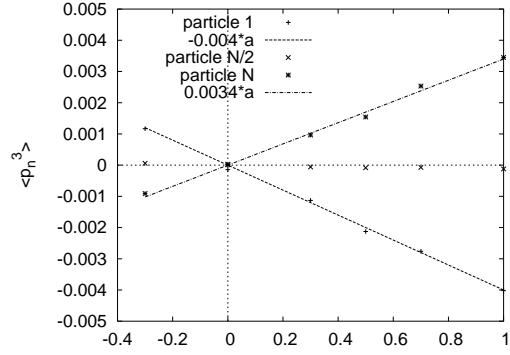}
 \end{center}
\caption{\small Variation of $\langle p_{1}^3\rangle$, $\langle p_{N/2}^3\rangle$ and 
$\langle p_{N}^3\rangle$ with parameter $a$.
The system and reservoir are the same as in Fig.\/\ref{f4} }
\label{f4x}
\end{figure}

Next, we consider the case of modifying the on-site potential. Then, 
the third-order term is inserted into the on-site potential of the 
$\phi^4$ model as 
\begin{equation}
 U(q)=\frac{1}{2}q^2+\frac{a}{3}q^3+\frac{1}{4}q^4,
\end{equation}
and the harmonic interaction potential $V(q)=q^2/2$ is used.
Figures \ref{f3} (a) and (b) show the deviation from the Maxwellian 
distribution for particles 8 and 16 in the system with $a=1.0$ of 
system size $N=16$. 
Asymmetric deviation is observed for the end particle 
but invisible in the middle of the chain.  
Figure \ref{f4} shows the third-order moments $\langle p_n^3\rangle$
and its variation with parameter $a$, which clearly indicates that only the
particles at the ends of the chain are affected by the asymmetry.  
Notice that even for the end particles the magnitude of moments is much 
smaller than that in the case of modifying interaction potential.
Namely, asymmetry in the on-site potential produces much smaller effects than
asymmetry in the interaction potential.
Figure \ref{f4x} shows
$\langle p_n^3\rangle$ vs $a$ for particles at the ends and one in the bulk. 
The moments change linearly with $a$ for particles at the ends but 
shows little changes for the particle in the bulk.
The linearity implies that this change is certainly caused by the
introduced asymmetry.  In the middle of the system, the effect of asymmetry
is too small to be observed.

\begin{figure}[htbp]
\begin{center}
\begin{tabular}{cc}
\includegraphics[width=7cm]{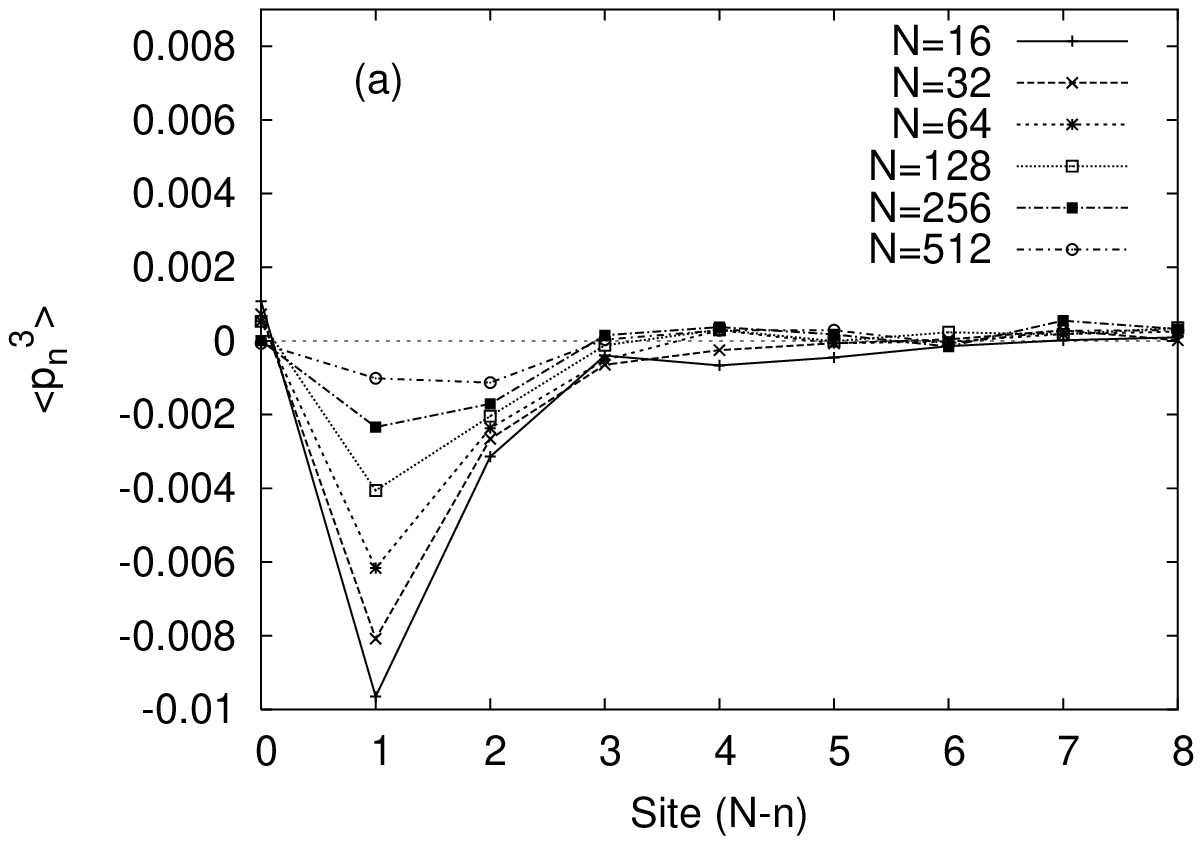}&
\includegraphics[width=7cm]{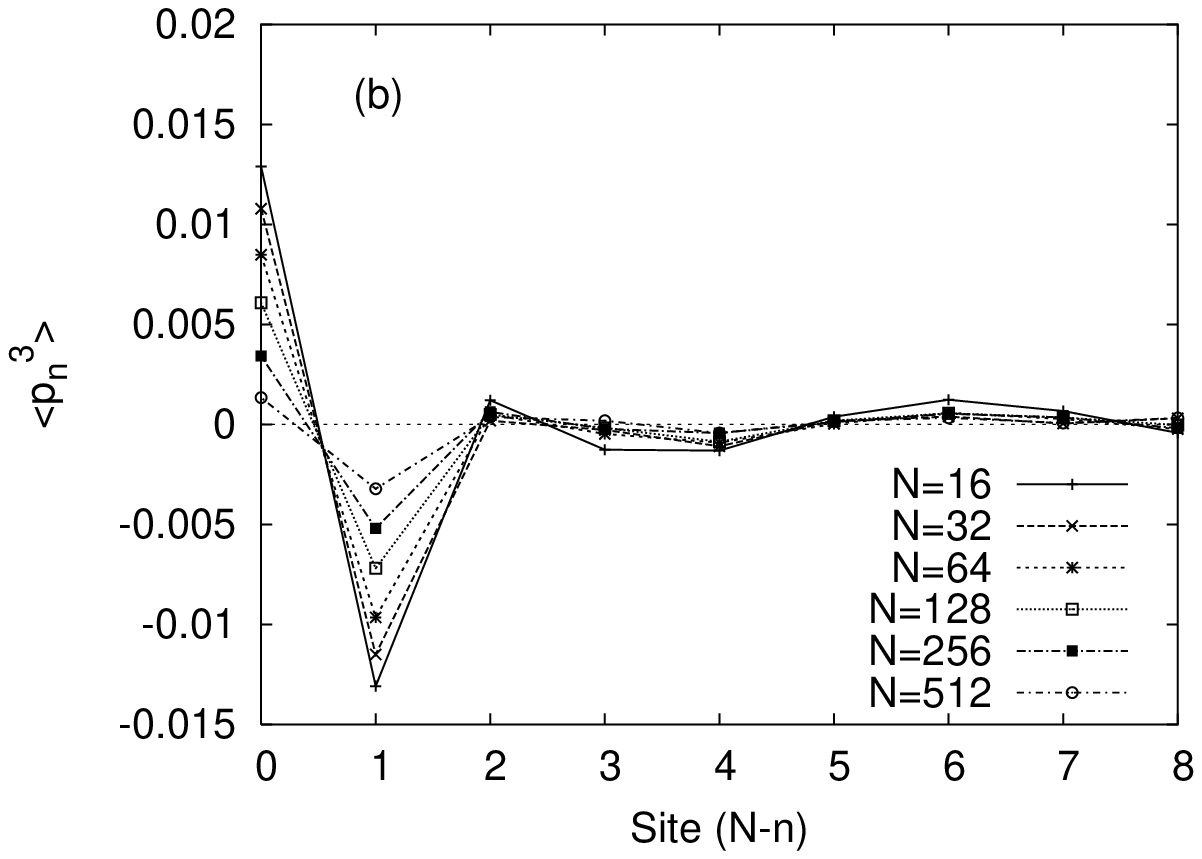}
\end{tabular}
 \end{center}
\caption{\small The third-order moments $\langle p_n^3\rangle$ near the right
 end in the system with the asymmetric on-site potential 
$U(q)=\frac{q^2}{2}+\frac{q^3}{3}+\frac{q^4}{4}$ and 
the interaction potential $V(q)=\frac{q^2}{2}+\frac{q^4}{4}$. 
The system size $N=16$, 32, 64, 128, 256 and 512.   Heat reservoirs are
(a) Langevin reservoirs and 
(b) Nose-Hoover reservoirs at $T_L=2.0$ and $T_R=1.0$.}
\label{f5x}
\end{figure}

\begin{figure}[htbp]
\begin{center}
\begin{tabular}{cc}
\includegraphics[width=7cm]{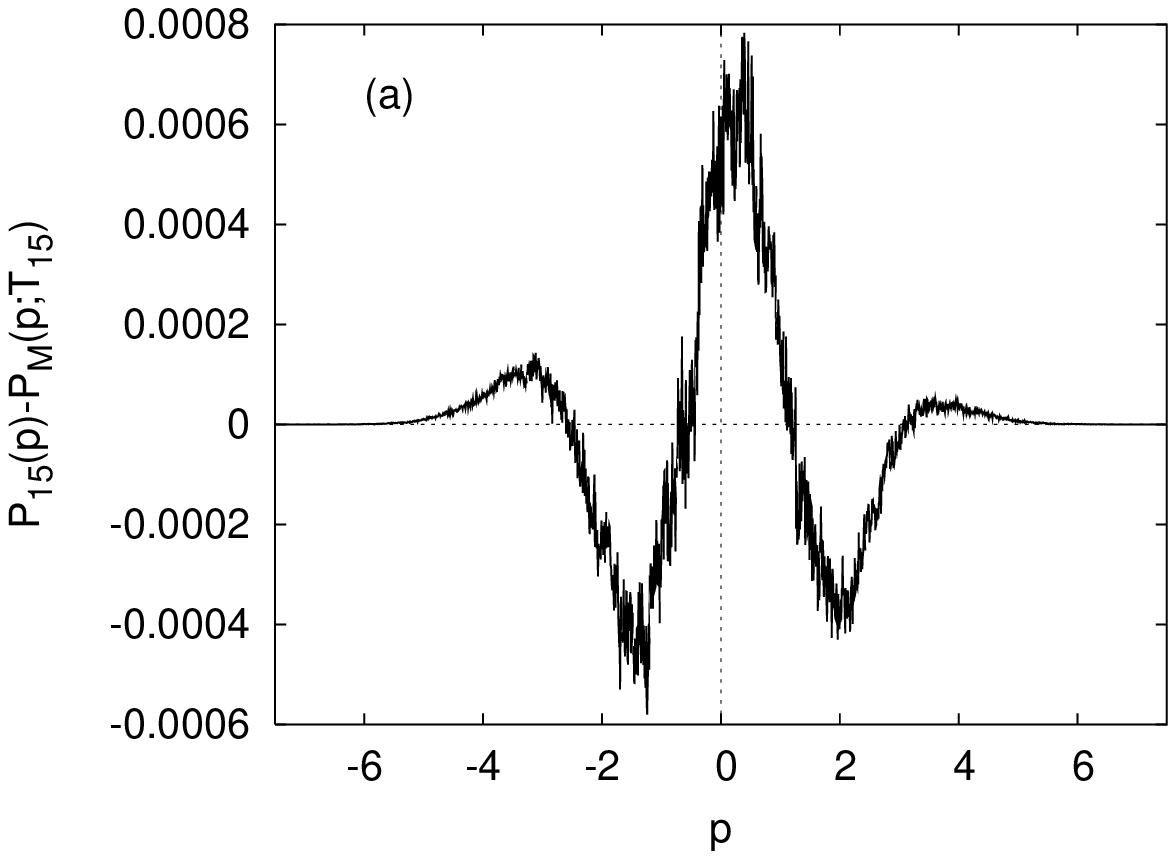}&
\includegraphics[width=7cm]{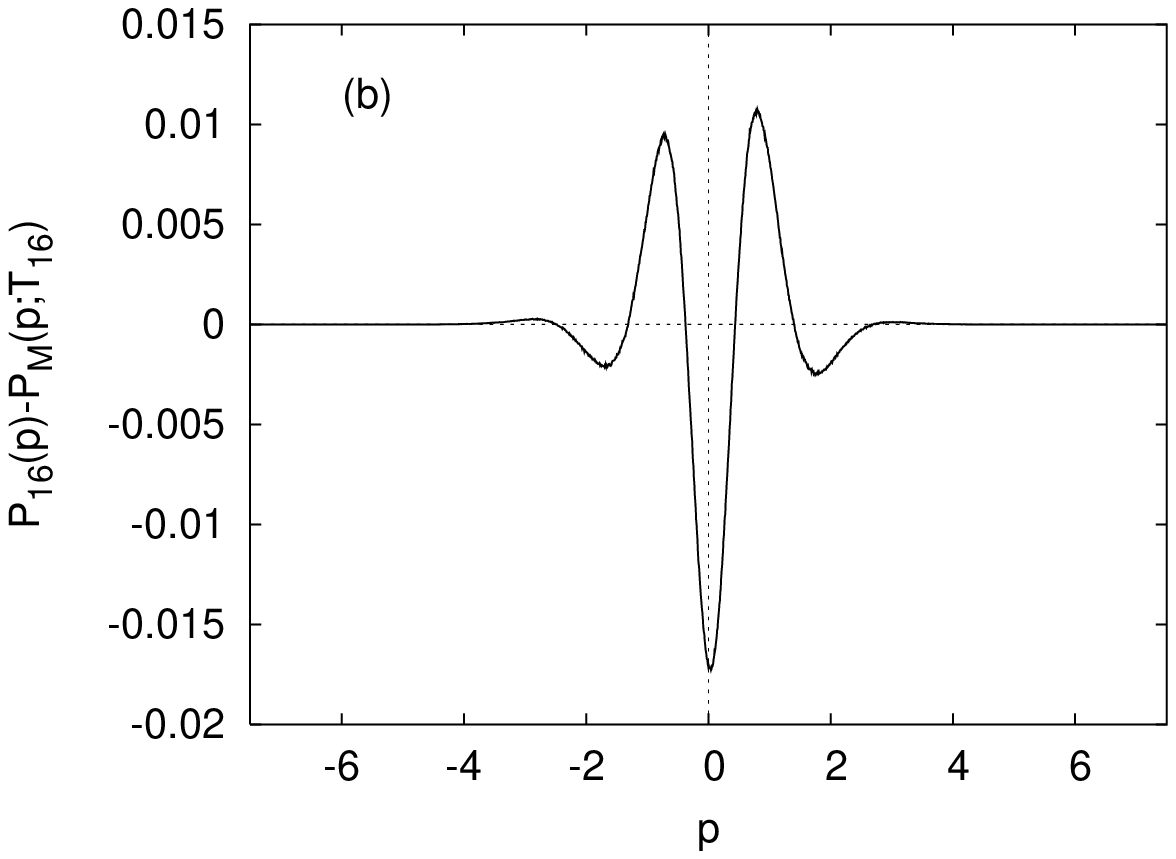}
\end{tabular}
\end{center}
\caption{\small  Deviations from the Maxwellian distribution in the system
of size 16 with the asymmetric on-site potential 
$U(q)=\frac{1}{2}q^2+\frac{1}{3}q^3+\frac{1}{4}q^4$ and the
nonlinear interaction potential $V(q)=\frac{1}{2}q^2+\frac{1}{4}q^4$.
(a) Particle 15 in the system with Langevin reservoirs.
(b) Particle 16 in the system with Nose-Hoover reservoirs. 
The reservoir temperatures are given as $T_L=2.0$ and $T_R=1.0$.
$10^{10}$ iterations are carried out for each of them.}
    \label{f4xx}
\end{figure}

When the interaction potential is not a harmonic potential but a nonlinear
one as
\begin{equation}
V(q)=\frac{1}{2}q^2+\frac{1}{4}q^4,
\label{inter} 
\end{equation}
the largest deviation of the momentum distribution is observed not for
particles $1$ and $N$ but particles $2$ and $N-1$.  However, the effect of
asymmetry is still confined to the vicinity of both the ends.  Figure 
\ref{f5x} (a) shows the third-order moments near the right end and their
system-size dependence.  As seen from this figure, the number of particles
affected by the asymmetry is not changed with the system size $N$.  If we
further replace the Langevin reservoirs with the Nose-Hoover ones, the
confinement of the asymmetric effects is not changed, as is shown in
Fig. \ref{f5x} (b).   A notable change is
that the magnitude of the moments becomes fairly large.
The shape of the deviation depends on the interaction potential
and the reservoirs.  Figure \ref{f4xx} (a)  shows the
deviation of the momentum distribution for particle $N-1$, which has the
largest moment in the system with the nonlinear interaction potential and 
Langevin reservoirs.  The deviation has similar shape with
Fig.\/ \ref{f3} (b).  However, the deviation changes its
sign if the reservoirs are replaced with the Nose-Hoover ones as seen in
Figure \ref{f4xx} (b), which represents the deviation for the rightmost
particle.   In this case, the deviation for particle $N-1$ and that for
particle $N$ have different signs.  
From these detailed observations, we conclude that the asymmetry effect is 
confined to the vicinity of the ends in the systems with an asymmetric on-site
potential and a symmetric interaction potential.   On the other hand, the
extent of the confinement and the shape of the deviation depend on the
details of the systems.

\begin{figure}[tb]
\begin{center}
\begin{tabular}{cc}
\includegraphics[width=7cm]{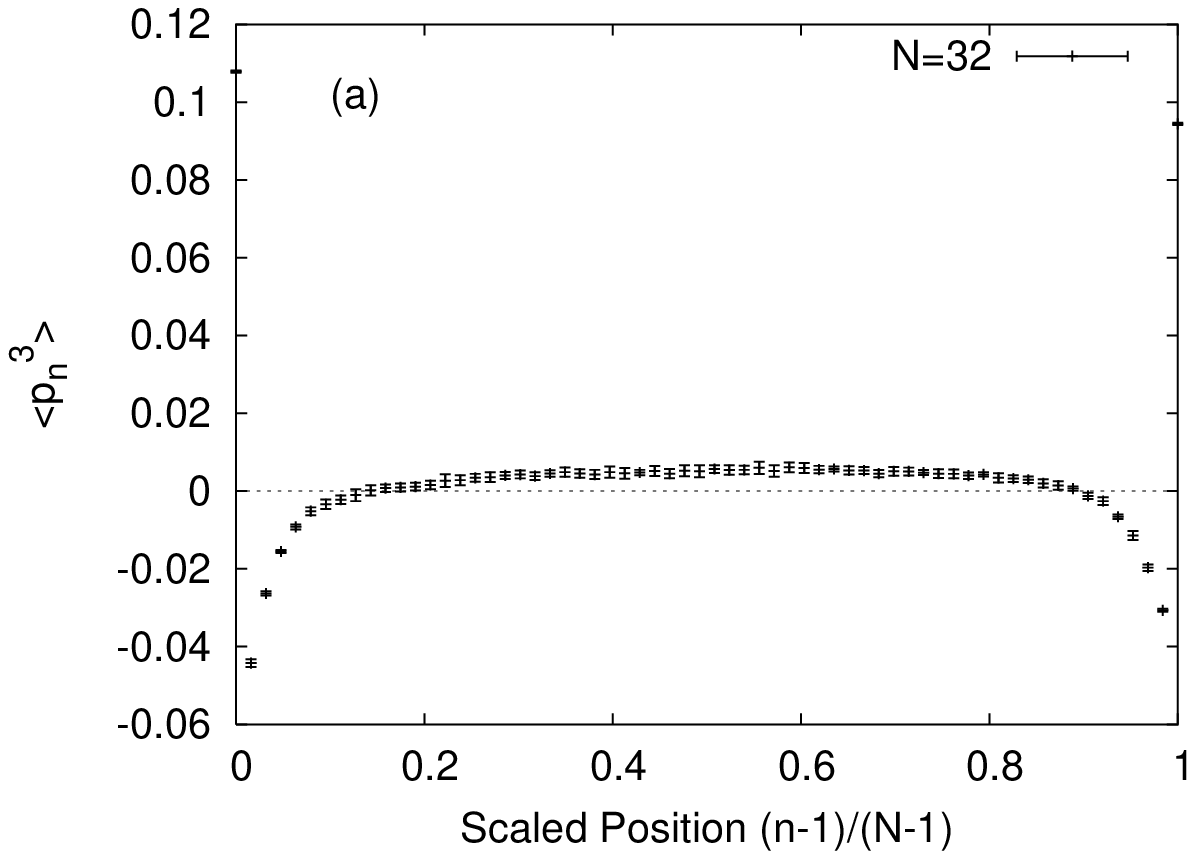}&
\includegraphics[width=7cm]{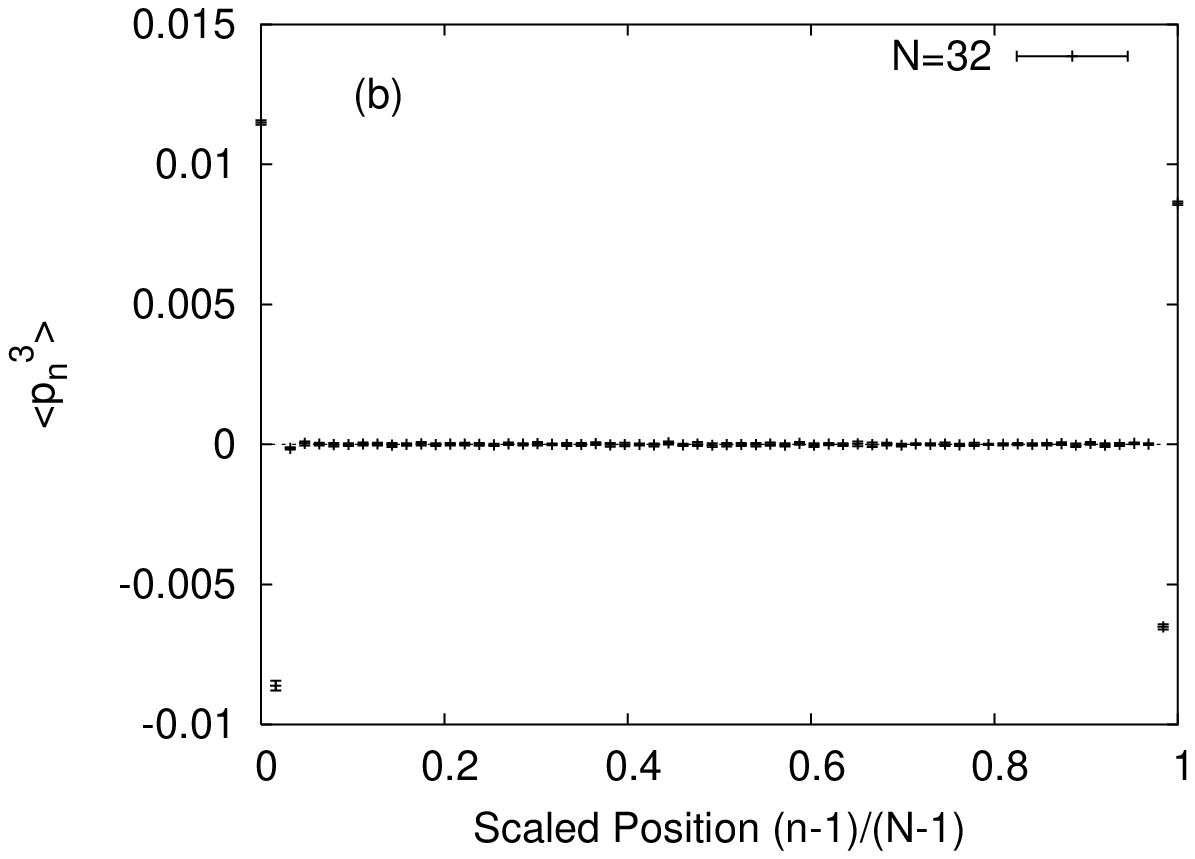}
\end{tabular}
\end{center}
\caption{\small The third-order moments $\langle p_n^3\rangle$ 
for (a) the FPU model and (b) the $\phi^4$ model when the thermal walls
at $T_L=2.0$ and $T_R=1.0$ are used.  The system size is $N=32$.
}
\label{f5}
\end{figure}

\begin{figure}[tb]
\begin{center}
\begin{tabular}{cc}
\includegraphics[width=7cm]{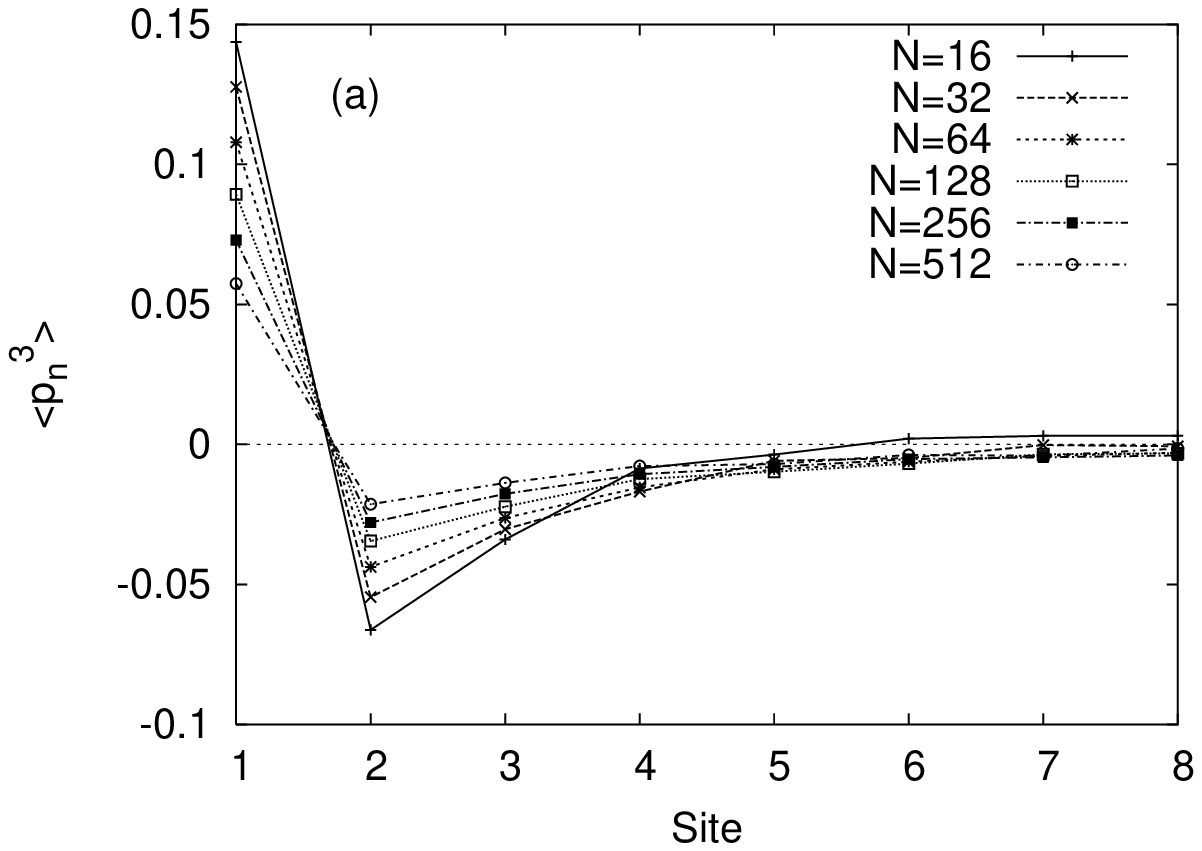}&
\includegraphics[width=7cm]{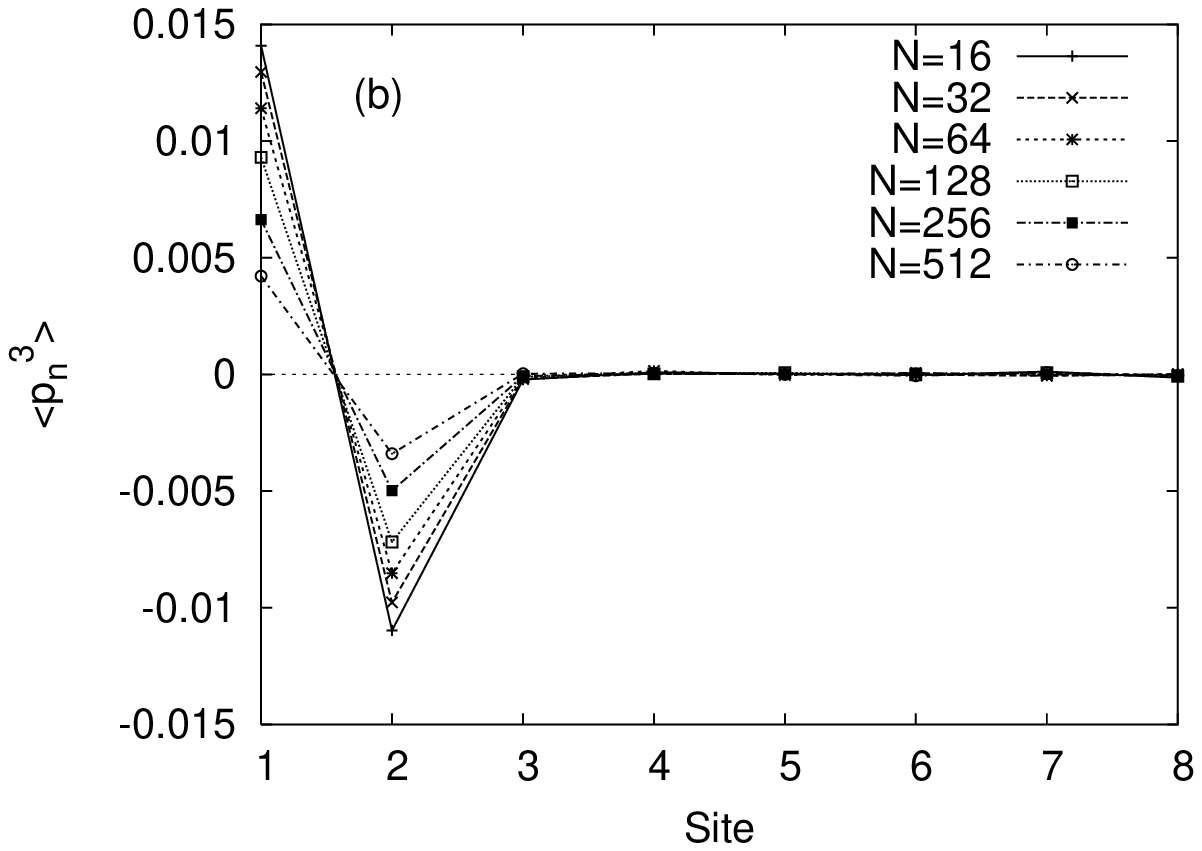}
\end{tabular}
\end{center}
\caption{\small The third-order moments $\langle p_n^3\rangle$ 
for (a) the FPU model and (b) the $\phi^4$ model 
when the thermal walls at $T_L=2.0$ and $T_R=1.0$ are used. 
The system size is $N=16, 32, 64, 128, 256, 512$.
}
\label{f6}
\end{figure}

Now we turn to the case where the heat reservoirs break the symmetry.
As mentioned in Sec.\/ 3, the Langevin reservoirs and the Nose-Hoover
reservoirs keep the symmetry but the thermal wall does not.  
Figure \ref{f5} compares the profiles of the third-order moments in (a) the
FPU model and (b) the $\phi^4$ model when the thermal walls are employed as 
heat reservoirs.  In the $\phi^4$ model, the influence of the
symmetry-breaking is virtually limited to only the four particles, $1$, $2$,
$N-1$ and $N$.  The moments almost vanish for the other particles. On the
contrary, in the FPU model, finite moments appear even in the central
part of the system, though the magnitude is much smaller than the case
of particles near the boundaries. Figures \ref{f6} (a) and (b) enlarge
the profile of moments near the ends, and show system-size dependence
for the FPU model and the $\phi^4$ model, respectively. In the $\phi^4$
model, the number of sites affected by the asymmetry does not change. In
the FPU model, the moments behave as if to decrease exponentially from the 
boundary into the bulk.  Interestingly, however, it actually goes 
beyond zero and finite moments are obtained in the central region.
Recall that the $\phi^4$ system has a bulk limit of the thermal conductivity,
while the FPU model does not.  Our result also implies that the FPU system
has no {\em bulk}. 

In the dilute gas system between two thermal walls at different temperatures, 
it is known that asymmetric deviations appear in the whole system\cite{Kim}. 
Thus the lattice system is totally different from the gas system in this
point.

\section{Summary and Discussion}
We have numerically studied the single-particle momentum distribution 
functions in one-dimensional lattice dynamical system in nonequilibrium 
steady states of heat conduction. 
We especially focus on the deviations from the Maxwellian distribution. 
This deviation reflects a symmetry of the system.  Namely,
if the system is invariant under the change of signs of the dynamical 
variables, the momentum distribution function must be even.
This symmetry can be broken by the interaction potential or the
on-site potential or the heat reservoir.  In the first case, the effect of
asymmetry extends to the whole system.  In the second case, the effect of
asymmetry is concentrated near the ends.  In the last case, 
we have found notable differences in systems with and without 
on-site potentials.  Namely, the effect of asymmetry is localized near the
ends in the systems with on-site potentials, whereas the effect
is extended to the center of the system without on-site potentials.   
This means there is no bulk limit in the system
without on-site potentials, which is consistent with the known results on
the convergence or divergence of thermal conductivity.

We need to develop a theoretical explanation for our present results and 
clarify a relation between the deviation and thermal conductivity.
Before studying that, it is necessary to specify what parameters are relevant 
to the deviations.  As we mentioned in Sec.\/ 3, deviation of the momentum
distribution is related to two-site correlations including heat flux.
Thus, heat flux and local temperature, which is the second-order moment of
the distribution, are considered to be important parameters.  It is a future
problem to clarify if they are sufficient to determine the deviations.

\section*{Acknowledgment}
The authors thank H. Hayakawa, M. M. Sano for valuable discussions
and helpful comments.  AU thanks H. Nishimori for continuous encouragement. 
This work is supported by the Grant-in-Aid for the
21st Century COE ''Center for Diversity and Universality in Physics''
from the Ministry of Education, Culture, Sports, Science and Technology
(MEXT) of Japan. The numerical computation in this work was carried out at 
Library \& Science Information Center, Osaka Prefecture University.

\end{document}